\documentclass{emulateapj}

\newcommand{\hi}{\mbox{\rm \ion{H}{1}}}

\newcommand{\htwo}{\mbox{\rm H$_2$}}

\newcommand{\halpha}{\mbox{\rm H$\alpha$}}

\newcommand{\aco}{\mbox{$\alpha_{\rm CO}$}}

\newcommand{\xcounits}{\mbox{${\rm M}_{\odot}$(K km s$^{-1}$ pc$^{2}$)$^{-1}$}}

\shorttitle{Unusually Luminous GMCs in the Outer Disk of M33}
\shortauthors{Bigiel et al.}

\begin{document}

\slugcomment{Accepted for publication in the Astrophysical Journal}
\title{Unusually Luminous Giant Molecular Clouds in the Outer Disk of M33}

\author{F.~Bigiel\altaffilmark{1} , A.~D.~Bolatto\altaffilmark{2}, A.~K.~Leroy\altaffilmark{3,7},
L.~Blitz\altaffilmark{1}, F.~Walter\altaffilmark{4}, E.~W.~Rosolowsky\altaffilmark{5},
L.~A.~Lopez\altaffilmark{6}, R.~L.~Plambeck\altaffilmark{1}}

\altaffiltext{1}{Department of Astronomy, Radio Astronomy Laboratory,
University of California, Berkeley, CA 94720, USA; bigiel@astro.berkeley.edu}
\altaffiltext{2}{Department of Astronomy, University of Maryland, 
College Park, MD 20742, USA}
\altaffiltext{3}{National Radio Astronomy Observatory, 520 Edgemont Road, Charlottesville,
VA 22903, USA}
\altaffiltext{4}{Max-Planck-Institut f{\"u}r Astronomie, K{\"o}nigstuhl 17,
69117 Heidelberg, Germany}
\altaffiltext{5}{Department of Physics and Astronomy, University of British Columbia Okanagan, 3333 University Way, Kelowna BC, V1V 1V7, Canada}
\altaffiltext{6}{Department of Astronomy and Astrophysics, University of California Santa Cruz,
1156 High Street, Santa Cruz, CA 95064, USA}
\altaffiltext{7} {Hubble Fellow}

\begin{abstract}
We use high spatial resolution ($\sim$7\,pc) CARMA observations to derive detailed properties for 8 giant molecular clouds (GMCs) at a galactocentric radius corresponding to approximately two CO scale lengths, or $\sim0.5$ optical radii (${\rm r_{25}}$), in the Local Group spiral galaxy M33. At this radius, molecular gas fraction, dust-to-gas ratio and metallicity are much lower than in the inner part of M33 or in a typical spiral galaxy. This allows us to probe the impact of environment on GMC properties by comparing our measurements to previous data from the inner disk of M33, the Milky Way and other nearby galaxies. The outer disk clouds roughly fall on the size-linewidth relation defined by extragalactic GMCs, but are slightly displaced from the luminosity-virial mass relation in the sense of having high CO luminosity compared to the inferred virial mass. This implies a different CO-to-\htwo\ conversion factor, which is on average a factor of two lower than the inner disk and the extragalactic average. We attribute this to significantly higher measured brightness temperatures of the outer disk clouds compared to the ancillary sample of GMCs, which is likely an effect of enhanced radiation levels due to massive star formation in the vicinity of our target field. Apart from brightness temperature, the properties we determine for the outer disk GMCs in M33 do not differ significantly from those of our comparison sample. In particular, the combined sample of inner and outer disk M33 clouds covers roughly the same range in size, linewidth, virial mass and CO luminosity than the sample of Milky Way GMCs. When compared to the inner disk clouds in M33, however, we find even the brightest outer disk clouds to be smaller than most of their inner disk counterparts. This may be due to incomplete sampling or a potentially steeper cloud mass function at larger radii.
\end{abstract}

\keywords{galaxies: individual (M33) --- galaxies: ISM --- ISM: clouds}

\section{Introduction}
\label{intro}

Star formation in normal, non-starburst galaxies is observed to occur predominantly in giant molecular clouds (GMCs), which are gravitationally bound structures of molecular gas. In order to understand star formation in galaxies, it is thus important to gain a thorough understanding of GMC properties and how they depend on environment within galaxies. In this paper we study the properties of GMCs in the outer disk of the local spiral galaxy M33 and study the impact of environment on GMC properties by comparing our measurements to those from the inner disk of M33 and from other local galaxies.

When GMCs were systematically observed in the Milky Way, their observable properties were found to obey a number of scaling relations \citep[`Larson laws',][]{larson81}: Size, line width, and CO luminosity are correlated in Galactic clouds \citep[e.g.,][]{solomon87,heyer09}. The original form of the size-line width relation ($R \propto \sigma^{0.5}$) in combination with the observation that molecular clouds are approximately in virial equilibrium also implies that their molecular gas surface densities are roughly constant. In the Milky Way, \citet{solomon87} find $\Sigma_{\rm H2}\approx170{\rm M}_{\odot}~{\rm pc}^{-2}$ from virial mass estimates using $^{12}$CO emission, though recently \citet{heyer09} found a much lower value of $\Sigma_{\rm H2}\approx50{\rm M}_{\odot}~{\rm pc}^{-2}$ with a scatter of $23{\rm M}_{\odot}~{\rm pc}^{-2}$ using more optically thin $^{13}$CO emission and assuming local thermodynamic equilibrium. The authors of the latter study point out, however, that this value could be as high as $\sim120 {\rm M}_{\odot}~{\rm pc}^{-2}$ due to abundance variations in outer cloud envelopes. 

With the advent of millimeter-wave interferometry, detailed studies of GMC properties became feasible for Local Group and relatively nearby galaxies. This includes observations in, e.g., M31 \citep{vogel87}, M33 \citep{wilson90,engargiola03} and other nearby galaxies \citep[see][and references therein]{walter01,walter02,bolatto08,blitz07}. These studies did not find large differences regarding GMC properties compared to the Milky Way.

The currently probed range in environmental parameter space is quite limited, however. Variations of cloud properties with environment might be expected though, as these properties might depend on, e.g., metallicity, interstellar radiation field or dust abundance \citep[e.g.][]{elmegreen89}. Few studies have addressed this issue in more extreme, in particular lower metallicity, environments and found only minor deviations (most notably in the Small Magellanic Cloud, SMC) from the previously established scaling relations \citep[e.g.,][referred to as B08 in the following]{walter01,walter02,leroy06,bolatto08}.

After extending parameter space by going to more metal poor galaxies, one of the next logical steps is to look for varying cloud properties within the same galaxy. Deep single dish observations revealed CO emission far out in or even beyond the optical disks of galaxies \citep[e.g.,][]{braine04,braine07}, including the Local Group spiral M33 \citep{gardan07,braine10,gratier10}. These observations probe molecular gas properties in a regime where conditions in the interstellar medium and thus GMC properties are expected to change, because of, e.g., changing metallicities or dust abundance. The proximity of M33 \citep[D$\approx$840\,kpc,][]{galleti04} and the availability of comprehensive, homogeneous measurements and consistently determined GMC properties across the central star forming disk \citep{rosolowsky03,rosolowsky07} makes this galaxy an ideal target for high resolution CO observations to probe GMC properties in the outer disk. 

\begin{figure*}
\plottwo{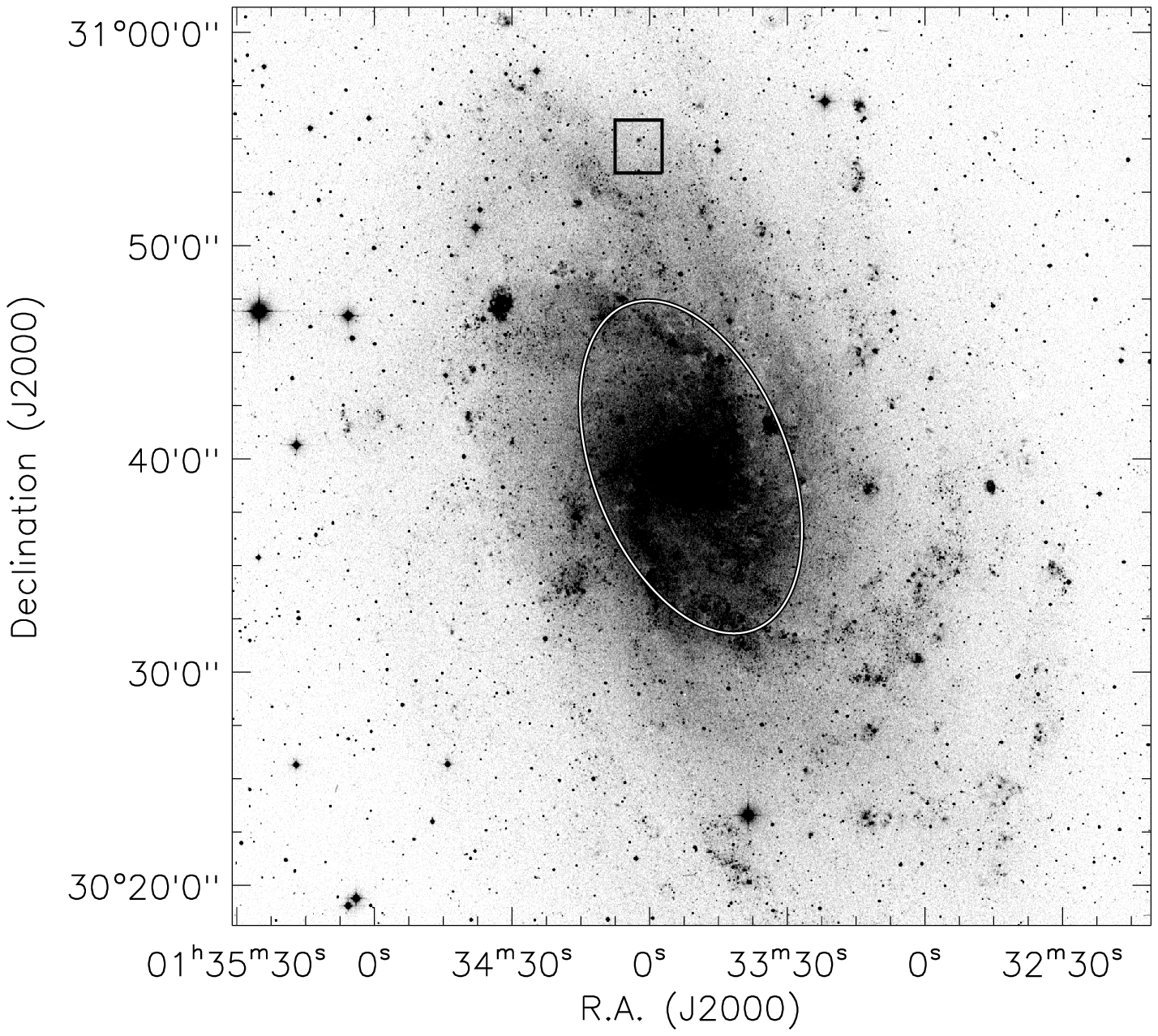}{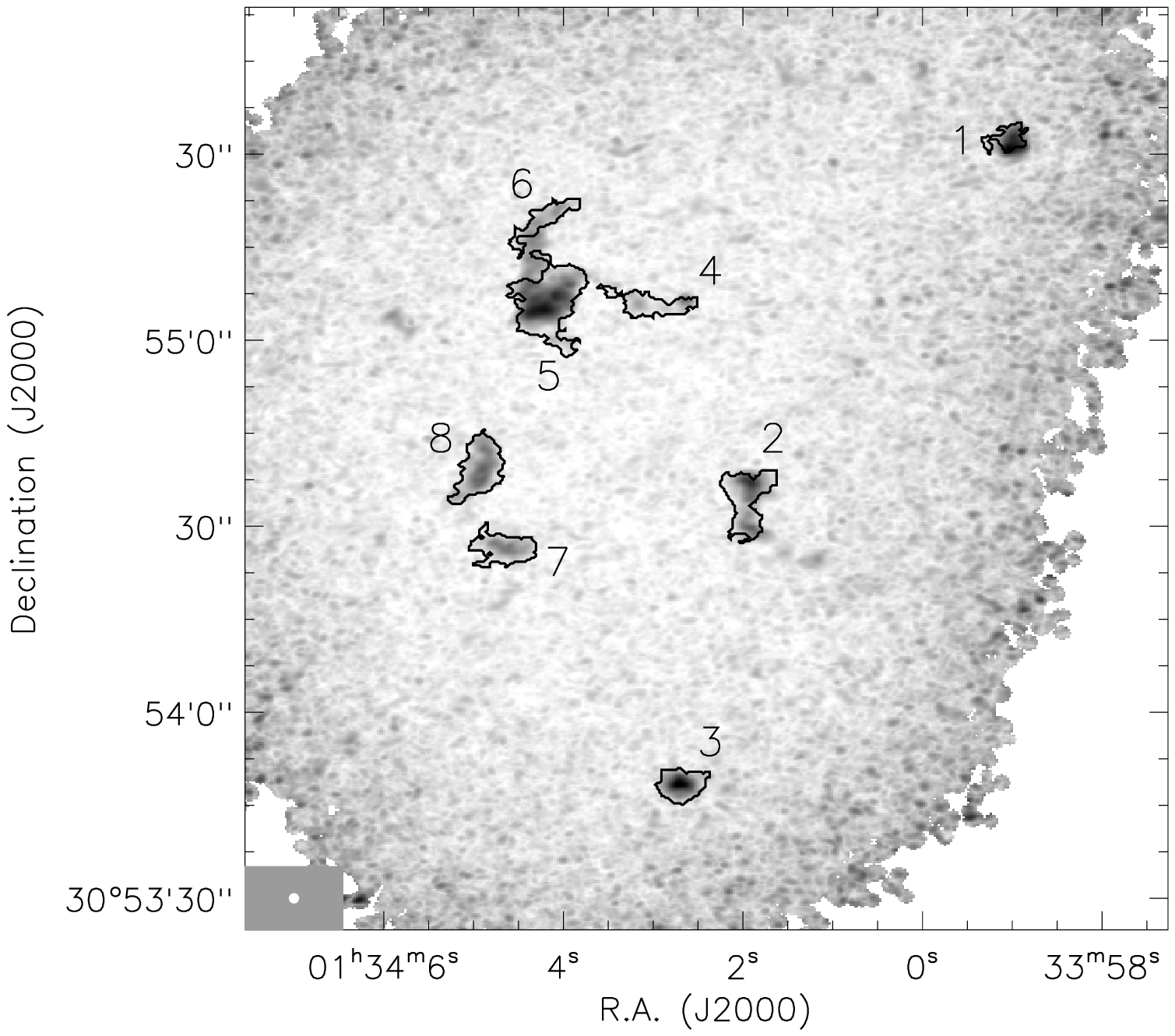}
\caption{{\em Left:} The location of the GMC complex analyzed in this paper (black rectangle; the size matches the peak intensity map in the right panel) relative to the distribution of young stars (DSS B-band image) in M33. The galactocentric distance of our target field corresponds to approximately two CO scale lengths (the white ellipse indicates one CO scale length or $r=$2\,kpc). It contains one of the furthest, large GMC complexes in the disk of M33 that is detected with high signal-to-noise in deep single dish observations by \citet{gardan07}. {\em Right:} $^{12}$CO(1-0) peak intensity map from CARMA of our target field. The size of the synthesized beam of $\sim1.7\arcsec$ is shown in the lower left corner. CPROPS identifies 8 GMCs in this field (highlighted by contours), for which, at $\sim7$\,pc resolution and with peak signal-to-noise ratios $\gtrsim10$, we can derive accurate cloud properties. The number next to each cloud identifies them in Table \ref{table1}, where we present the derived properties.}
\label{fig1}
\end{figure*}

In this paper we present CO observations from the Combined Array for Research in Millimeter Wave Astronomy (CARMA) of 8 GMCs. The galactocentric distance of these clouds corresponds to approximately two scale lengths of the molecular gas, i.e., the length over which the azimuthally averaged CO emission declines by a factor $e^2$, or $\sim0.5{\rm r_{25}}$, where r$_{25}$ is the isophotal radius corresponding to 25 B-band magnitudes per square arcsecond. Figure \ref{fig1} illustrates the location of our targeted field and shows the CO peak intensity map we obtain. We introduce our observations which we use to measure cloud properties for 8 GMCs in Section \ref{data} and discuss the expected environmental variations (molecular gas fraction, dust-to-gas ratio, metallicity) within M33 with respect to our target field in Section \ref{environ}. We compare our measurements to those from the inner disk of M33, other nearby galaxies and the Milky Way in Section \ref{cloud-props}.

\section{Data \& Methodology}
\label{data}

Our CARMA observations were targeted towards one of the outermost detected molecular gas complexes from the single dish observations of \citet[][compare Figure \ref{fig1}]{gardan07}. We note that recently \citet{braine10} have obtained deeper single dish data and detected CO emission even further out in the disk of M33. We observed the target position in the CO(1-0) transition using a 19-point mosaic in D configuration between July and August 2008 and in C configuration between October and November 2009 under mostly good 3mm weather conditions. The total observing time was $\sim$37\,hr and $\sim$30\,hr in D and C configuration, respectively. The synthesized beam size is $\sim1.7\arcsec$, corresponding to $\sim7$\,pc at the distance of M33. We tuned the 3mm receivers to the doppler-shifted CO(1-0) rest frequency in the upper sideband and placed two $\sim31$\,Mhz wide bands across the line. This yields an effective bandwidth of $\sim$155\,km\,s$^{-1}$ and a velocity resolution $\delta {\rm v_{channel}}=1.27$\,km\,s$^{-1}$. For phase and amplitude calibration, 0205+322 was observed every 20 minutes. Fluxes were bootstrapped from Uranus, Neptune or MWC349 and 3C454.3 or 3C84 were observed for bandpass calibration. Radio and optical pointing was performed every 2\,hr. We estimate the resulting calibration to be accurate within $\sim15\%$. Data reduction was performed in MIRIAD. Figure \ref{fig1} shows the target field with respect to the young stars (left panel) and the CO(1-0) peak intensity map from our observations (right panel).

We identify GMCs and measure their properties using the CPROPS package \citep{rosolowsky06}. We define GMCs as contiguous regions of high signal-to-noise emission. No further decomposition appears necessary in this dataset. CPROPS uses moment methods to measure radius, line width, luminosity, and peak temperature for each cloud. It estimates the associated uncertainties using boostrapping techniques and attempts to correct each measurement for biases due to finite sensitivity and limited spectral and angular resolution. A key point for this study is that B08 have used exactly the same approach to measure GMC properties from a large sample of nearby galaxies, including the inner disk of M33. This gives us a well-controlled point of comparison.

Table \ref{table1} lists the derived properties for the 8 GMCs in our field. We note that more faint CO features are visible in Figure \ref{fig1}, but the signal-to-noise ratio in these features is too low for reliable cloud property estimates.

\begin{deluxetable*}{rrrrrrrrrr}
\tablecaption{GMC Properties} \tablehead{
\colhead{Number\tablenotemark{a}} & \colhead{R.A.} & \colhead{Dec.} & \colhead{${\rm v_{lsr}}$} & \colhead{R\tablenotemark{b}} & \colhead{$\sigma$\tablenotemark{b}} & \colhead{${\rm M_{lum}}$\tablenotemark{b}} &
\colhead{${\rm M_{vir}}$\tablenotemark{b}} & \colhead{${\rm T_{B}}$\tablenotemark{c}} & \colhead{${\rm \Sigma_{GMC}}$\tablenotemark{d}} \\
\colhead{} & \colhead{[J2000]} & \colhead{[J2000]} & \colhead{[km\,s$^{-1}$]} & \colhead{[pc]} & \colhead{[km\,s$^{-1}$]} & \colhead{[$10^{3}{\rm M}_{\odot}$]} &
\colhead{[$10^{3}{\rm M}_{\odot}$]} & \colhead{[K]} & \colhead{[${\rm M}_{\odot}~{\rm pc}^{-2}$]}  }
\startdata
  1&      01 33 59.0&      30 55 32.5&     -254.6&      10.1$\pm$1.9&      1.6$\pm$0.5&      47$\pm$4&      27$\pm$18&      10.4$\pm$0.6&      84$\pm$64\\
  2&      01 34 01.9&      30 54 34.2&     -257.4&      18.3$\pm$1.4&      1.6$\pm$0.3&     105$\pm$7&      46$\pm$20&      8.7$\pm$0.6&      44$\pm$20\\
  3&      01 34 02.7&      30 53 48.3&     -258.1&      12.3$\pm$1.4&      1.5$\pm$0.3&      76$\pm$9&      30$\pm$14&      11.0$\pm$0.6&     64$\pm$32\\
  4&      01 34 03.0&      30 55 05.9 &    -256.5&      14.8$\pm$2.1&      2.1$\pm$0.5&      62$\pm$9&      70$\pm$39&      6.5$\pm$0.6&      102$\pm$64\\
  5&      01 34 04.2&      30 55 06.5 &    -256.8&      22.9$\pm$1.6&      2.0$\pm$0.3&     233$\pm$9&      91$\pm$29&      9.9$\pm$0.6&      55$\pm$19\\
  6&      01 34 04.2&      30 55 19.3&     -256.6&      13.4$\pm$1.8&      2.7$\pm$0.5&      60$\pm$8&      104$\pm$48&      5.7$\pm$0.6&     185$\pm$100\\
  7&      01 34 04.7&      30 54 26.4&     -255.2&      16.1$\pm$1.3&      2.3$\pm$0.4&      91$\pm$9&      90$\pm$36&      6.1$\pm$0.6&      111$\pm$47\\
  8&      01 34 04.9&      30 54 39.2&     -265.2&      17.0$\pm$1.5&      0.9$\pm$0.1&      84$\pm$5&      14$\pm$2&      7.0$\pm$0.6&       15$\pm$4
\enddata
\tablenotetext{a}{The numbers refer to the individual clouds in the right panel of Figure \ref{fig1}.}
\tablenotetext{b}{Uncertainties are estimated from bootstrapping in CPROPS.}
\tablenotetext{c}{The uncertainty on ${\rm T_{B}}$ is estimated from the RMS scatter of the noise in the data cube.}
\tablenotetext{d}{${\rm \Sigma_{GMC}}$ is derived from the virial masses and radii in this table. Uncertainties are estimated from error propagation.}
\label{table1}
\end{deluxetable*}

\section{Environment}
\label{environ}

Previous interferometric observations of individual GMCs in M33 focussed on GMC properties in the inner disk\footnote{We compare our measurements to the M33 GMC sample from  \citet{bolatto08}, which has a median galactocentric distance of less than one CO scalelength.} of M33 \citep[e.g.,][]{engargiola03,rosolowsky03,rosolowsky07}. In the following we assess the distinguishing characteristics of the ISM at larger radii in M33 and how this may impact GMC properties.

Numerous studies have compared the radial behavior of atomic gas, molecular gas and various star formation tracers in M33 \citep[e.g.,][]{engargiola03,corbelli03,heyer04,gardan07}. While the \hi\ is found to remain relatively constant across the disk of M33, these studies derive a range of scale lengths between $\sim1.4-2.5$\,kpc for the CO emission. We note that we adopt a value of $2$\,kpc for the CO scale length throughout this paper. From the different behavior of \hi\ and CO emission, it is clear that the molecular gas fraction $\Sigma_{\rm H2} / \Sigma_{\rm HI}$ decreases as a function of radius. Most recently, \citet{gratier10} measured this fraction and find that while the gas is mostly molecular in the center, the molecular gas fraction decreases to only $\sim10-20\%$ at our target position at r$\approx4$\,kpc (adjusted to our adopted Galactic CO-to-\htwo\ conversion factor of  X$_{\rm CO}=2\times10^{20}\,{\rm cm^{-2}\, (K\,km\,s^{-1})^{-1}}$).

The stellar surface density, $\Sigma_{\rm star}$, in M33 is low compared to other, larger spirals \citep[e.g.,][]{braine10}. As $\Sigma_{\rm star}$ is important for the hydrostatic gas pressure, which was found to correlate well with the molecular gas fraction in nearby galaxies \citep[e.g.,][]{wong02}, this may be an explanation for the unusually low \htwo\ fraction. Another factor that may impact the \htwo\ fraction is the amount of dust in the ISM, as dust grains serve as sites of \htwo\ formation and shield the molecular hydrogen from photodissociation. \citet{gratier10} derive 24, 70 and 160\,$\micron$ scale lengths of $\sim1.4$, $\sim1.5$ and $\sim1.8$\,kpc, respectively, all of which are quite similar to the CO scale length in M33. In combination with the flat \hi\ profile, this indicates 1) that the dust-to-gas ratio is significantly lower ($\sim25\%$) at r$\approx$4\,kpc compared to the inner disk and 2) that the dust-to-gas ratio may play an important role in setting the molecular gas fraction. 

Because the dust-to-gas ratio also scales with metallicity, one might also expect a lower metallicity in our target region. Even though earlier measurements suggested a quite significant metallicity gradient across M33 \citep[up to $\sim-0.1$\,dex\,kpc$^{-1}$,][]{vilchez88,garnett97}, recent measurements provide evidence for a much shallower gradient of $\lesssim-0.03$\,dex\,kpc$^{-1}$ \citep{crockett06,rosolowsky08}. Compared to the center, this implies a metallicity that is lower by $\sim30\%$ at $r\approx4$\,kpc, i.e., Z$\approx8.24$, adopting a central value of 8.36 from \citet{rosolowsky08}. This metallicity corresponds to only $\sim30\%$ of the Galactic average.

A lower metallicity, i.e., lower carbon and oxygen abundance, and dust-to-gas ratio in the ISM are likely to affect GMC properties, because they lead to lower \htwo\ formation rates and less effective shielding from UV radiation \citep[e.g.,][]{mckee89,elmegreen89,maloney88}. Thus, one might expect GMCs to have higher (surface) densities in such environments and possibly to consist of smaller CO-bright cores embedded in CO-dark \htwo\ envelopes, leading to a higher CO-to-\htwo\ conversion factor \citep[e.g.,][]{madden97,leroy07,elmegreen89}. Our new measurements in combination with ancillary data allow us to test these expectations in M33 and to extend such studies into a barely probed regime of ISM parameter space.

\section{GMC Properties and Comparison to other Measurements}
\label{cloud-props}

\begin{figure*}
\plotone{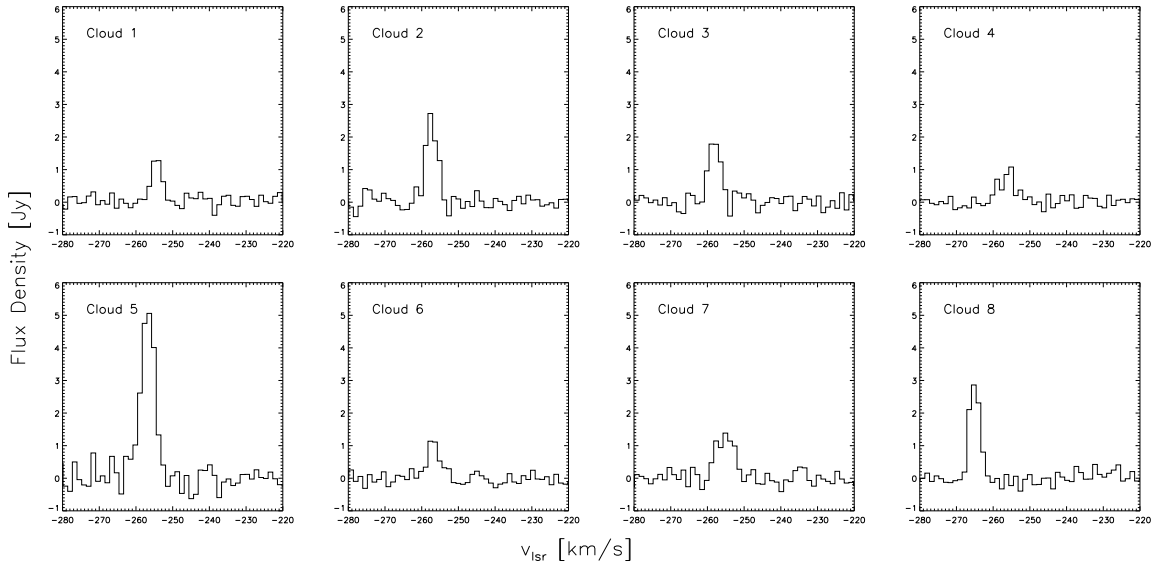}
\caption{Spectra of integrated emission within the isophotal contours in the right panel of Figure \ref{fig1} for each cloud in each channel. All clouds are clearly detected in several channels. The cloud numbers correspond to the numbers in Table \ref{table1}, listing the derived cloud properties.}
\label{fig3}
\end{figure*}

Figure \ref{fig3} shows spectra for the 8 identified GMCs. These clouds have relatively narrow spectra and the derived line widths are notably smaller than those from clouds from the inner disk of M33 (see discussion for Figure \ref{fig2} below). This agrees qualitatively with the finding of \citet{braine10}, though their spatial resolution is much lower, which may lead to overestimating line widths by averaging over multiple clouds. Most of our clouds have peak flux densities between $\sim1-3$\,Jy, with the exception of cloud 5 ($\sim5$\,Jy), which is the biggest and most massive GMC in our sample (though note the similarly small inferred velocity dispersion of $\sim2$\,km\,s$^{-1}$ compared to the other clouds in Table \ref{table1}).

B08 have assembled and partly reanalyzed CO observations of 3 disk galaxies (the Milky Way and the Local Group spirals M31 and M33) and 11 Local Group and nearby dwarf galaxies. Because in this paper we follow the same methodology, we can directly compare the properties of the B08 compilation of GMCs to our outer disk GMCs in M33. Comparing the peak brightness temperatures, we find a mean value for the new GMCs that is significantly higher than in the B08 ensemble: ${\rm T_{B,new}=8.2\,K}$ with a $1\sigma$ rms scatter of 3.3\,K and ${\rm T_{B,B08}=2.3\,K}$ with a scatter of 1.2\,K. For M33 specifically, all of the outer disk clouds have higher ${\rm T_{B}}$ than all but one of the inner disk clouds. ${\rm T_{B}}$ depends mainly on the beam filling fraction and the excitation temperature (it also depends on optical depth, but the $^{12}$CO(1-0) emission is most likely optically thick in all cases). In B08, typical brightness temperatures within individual galaxies rarely exceed 3.5\,K, even for cases where GMCs were observed at high spatial resolution (comparable to this study) so that GMCs were comfortably resolved. This also includes the previous, inner disk M33 observations. At least for these higher resolution data, the equally high beam filling fractions $\gtrsim1$ argue for higher kinetic temperatures in the molecular gas (under the assumption of uniform brightness distributions within GMCs) to explain the higher peak brightness temperatures we measure. A possible explanation might be the potentially elevated radiation field due to massive star formation in our target field.

\begin{figure}
\plotone{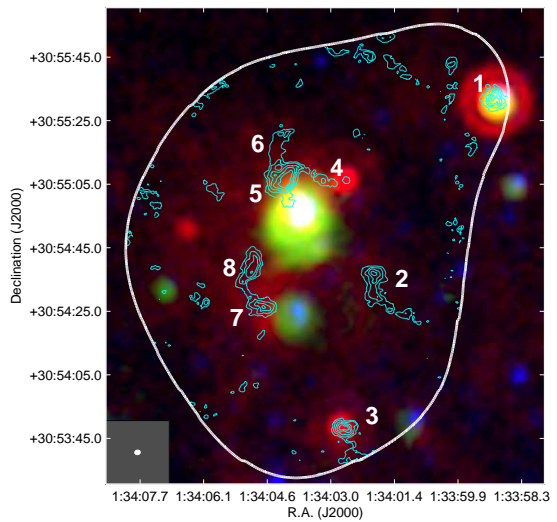}
\caption{Comparison of 24\micron\ (red), far UV (blue) and \halpha\ emission (green) in our target region. The full sensitivity field-of-view is indicated with a white line and integrated intensity contours of our CO observations are shown in cyan. The contours are running from 0.25-1.2\,Jy\,beam$^{-1}$\,km\,s$^{-1}$ in steps of 0.2\,Jy\,beam$^{-1}$\,km\,s$^{-1}$. The size of the synthesized beam is shown in the lower left corner. The cloud numbers are identical to those in Figure \ref{fig1} and refer to the cloud properties in Table \ref{table1}. The figure shows several regions of intense 24\micron\ and \halpha\ emission (both indicating current star formation) in the immediate vicinity of the detected GMCs.}  
\label{fig4}
\end{figure}

\begin{figure*}
\plotone{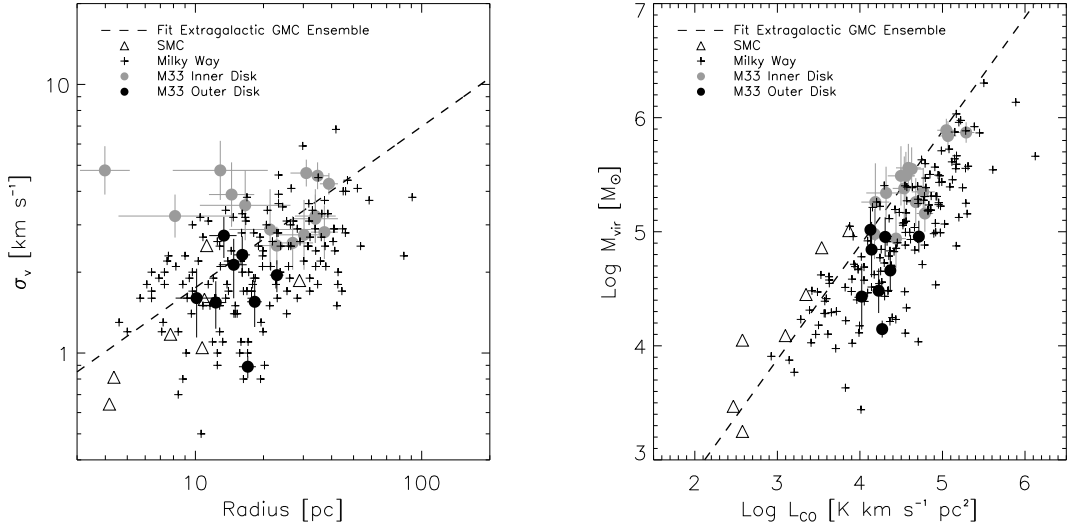}
\caption{Size-line width (left panel) and CO luminosity-virial mass (right panel) plots for the new M33 clouds (black points, this study) as well as for a number of other datasets for comparison: inner disk M33 \citep[gray points,][B08]{rosolowsky03} and SMC clouds (triangles, B08), Milky Way GMCs \citep[black crosses,][B08]{solomon87,heyer09} and a fit to an ensemble of extragalactic GMCs (dashed line, B08). The new clouds extend the distribution of the previously known clouds in the inner disk of M33 toward smaller sizes and line widths as well as toward smaller virial masses and lower CO luminosities. The combined M33 inner and outer disk sample covers about the same phase space in both plots as the Milky Way clouds.}
\label{fig2}
\end{figure*}

We illustrate this in Figure \ref{fig4}, where we show 24\micron\ (red), far UV (blue) and \halpha\ emission \citep[][green]{hoopes00} in our target region. Overlaid are the CO integrated intensity contours in cyan and our full sensitivity field-of-view (white line). From this figure it becomes clear that our GMCs are in the immediate vicinity of several star forming regions. In particular the two clouds with the highest peak brightness temperatures, 1 and 3, are directly associated with \halpha\ and 24\micron\ sources. The combination of presumably low dust-to-gas ratios, low gas columns and intense star formation supports our interpretation that the molecular gas is heated to high temperatures leading to the unusually high peak brightness temperatures we measure.

In Figure \ref{fig2} we analyze scaling relations for our GMCs: we plot the size-linewidth relation in the left panel and the relation between luminosity and virial mass in the right panel. In both panels, the dashed line shows the best fit to the cloud ensemble from B08 and the gray points represent GMCs from the inner disk of M33 from \citet{rosolowsky03}, but as reanalyzed in B08 using CPROPS. We also add clouds from the Small Magellanic Cloud (SMC; triangles), the most metal-poor system in the B08 sample (${\rm Z_{SMC}\approx0.2\,Z_{MW}}$). Black crosses represent Milky Way GMCs from \citet{heyer09} (who reanalyzed the \citet{solomon87} sample), which includes a number of small clouds quite similar in size to our sample. We reiterate that they use a somewhat different  methodology and a different tracer ($^{13}$CO emission) compared to the other studies (see Section \ref{intro}). The black points show the new M33 outer disk GMCs from Table \ref{table1}. The error bars show estimates of the uncertainties from bootstrapping in CPROPS.

The left panel shows that 1) the new GMCs roughly fall on the B08 fit, but that 2) the data points are offset compared to the majority of inner disk clouds in M33. Due to limited sensitivity and resolution in the extragalactic measurements, however, small GMCs are not always readily observable. For the Milky Way, however, where observations do have sufficient resolution and sensitivity, the GMCs overlap the combined inner and outer disk M33 points, indicating that the outer disk GMCs do not show dramatically different properties compared to what has been measured in the Galaxy. Nonetheless, most of the clouds we observe at $\sim0.5{\rm r_{25}}$ are smaller and tend to show lower velocity dispersions than most of the clouds in the inner disk of M33. This might hint at either a steeper cloud mass function or incomplete sampling of the latter.

We compute surface densities for our sample via ${\rm M_{vir} (\pi\,r^{2})^{-1}}$ and derive a mean surface density of ${\rm 82\,M_{\odot}~pc^{-2}}$ with a $1\sigma$ rms scatter of ${\rm 43\,M_{\odot}~pc^{-2}}$. This value is well within the range of values quoted in the literature, $50{\rm M}_{\odot}~{\rm pc}^{-2}\lesssim\Sigma_{\rm H2}\lesssim170{\rm M}_{\odot}~{\rm pc}^{-2}$, including Galactic clouds (compare Section \ref{intro}) or the extragalactic sample in B08. Thus, we find quite typical surface densities rather than much larger values as might have been expected (Section \ref{environ}).

The CO luminosity-virial mass plot in the right panel shows that the outer disk M33 GMCs seem to be an extension of the distribution of inner disk clouds toward lower CO luminosities and virial masses. As in the left panel, the combined M33 sample overlaps the Milky Way ensemble, indicating a similar range of virial and luminous GMC masses in both spiral galaxies. The new M33 clouds appear to fall slightly below the ensemble fit of B08, however. Because the normalization (intercept) of the power-law fit relating virial mass to CO luminosity yields the CO-to-\htwo\ conversion factor, \aco, the offset of our data points from the B08 fit implies a different average \aco\ (assuming identical power law slopes).

In order to assess this quantitatively, we compute \aco\ from the virial and luminous masses for each of the inner and outer disk clouds in M33 and find mean values of $6.8\pm0.8\,\xcounits$ and $3.3\pm0.8\,\xcounits$, respectively (the quoted errors represent the 1$\sigma$ uncertainty in the mean). B08 derive $7.6^{+3.9}_{-2.6}\,\xcounits$ for the extragalactic ensemble (the errors represent the scatter in the data), which is in good agreement with the M33 inner disk value. We test the significance of the difference between the M33 inner and outer disk means with a Student's t-test, which yields a probability of $\sim1\%$ for the null hypothesis that both means are equal. This implies that the average CO-to-\htwo\ conversion factor for the outer disk GMCs is in fact smaller (by about a factor of two) than the extragalactic and the M33 inner disk average. This finding is contrary to the expectations based on environmental conditions, which suggested a somewhat higher conversion factor (Section \ref{environ}).

One factor that could influence \aco\ (and is particularly relevant in outer disks) is a decreasing fraction of CO emitting \htwo\ in the more dust and metal-poor environment at larger radii (see Section \ref{environ}). We therefore compare our measurements to complementary \aco\ estimates in M33 from \citet{leroy10}. They estimate \aco\ from dust modeling (thus tracing the entire \htwo\ distribution under the assumption that gas and dust are well mixed) and find values of $\sim6.3\,\xcounits$ and $\sim4.7\,\xcounits$ for the inner and outer part of M33, respectively. These numbers are in good agreement with our virial mass estimates, which argues against a significant amount of CO-dark \htwo\ at large radii. We note that because the \citet{leroy10} values are averages over a large area, the application of these values to our specific region does not rule out conclusively unaccounted for \htwo. 

Taken at face value, however, the lower average conversion factor we measure for the outer disk clouds could be interpreted as a higher fractional CO abundance, i.e., more ``CO per \htwo''. This is highly unlikely, however, given the much lower dust-to-gas ratios and metallicities at larger radii in M33 (compare Section \ref{environ}). On the other hand, \aco\ scales inversely with the brightness temperature: $\aco\propto{\rm T_{B}}^{-1}$ \citep{dickman86,maloney88}. Thus, the more likely explanation seems to be the systematically higher ${\rm T_{B}}$ for the outer disk GMCs, which we interpreted above as higher kinetic gas temperatures due to elevated radiation levels from nearby star formation. The higher temperatures could lead to higher excitation of the CO molecules \citep[e.g.,][]{weiss01}, thus lowering the measured \aco.

With the resolution and sensitivity of CARMA we are able to measure the properties of 8 GMCs in the heavily \hi-dominated outer part of M33. Despite an environment very distinct from a normal spiral galaxy (low molecular gas fraction, stellar surface density and dust-to-gas ratio), the clouds we observe show generally similar properties compared to GMCs in the Milky Way or other nearby galaxies. The main difference is that the gas appears to be hotter, with excitation temperatures between $\sim6-11$\,K, which is likely to be the responsible mechanism for a lower inferred CO-to-\htwo\ conversion factor. This difference appears mostly attributable to heating by massive star formation coincident or adjacent to the GMCs.

\acknowledgments

F.B. acknowledges support from NSF grant AST-0838258. A.B. acknowledges partial support from NSF grant AST-0838178. Support for A.L. was provided by NASA through Hubble Fellowship grant HST-HF-51258.01-A awarded by the Space Telescope Science Institute, which is operated by the Association of Universities for Research in Astronomy, Inc., for NASA, under contract NAS 5-26555. 
E.R. is supported by a Discovery Grant from NSERC of Cananda. Support for CARMA construction was derived from the states of California, Illinois, and Maryland, the James S. McDonnell Foundation, the Gordon and Betty Moore Foundation, the Kenneth T. and Eileen L. Norris Foundation, the University of Chicago, the Associates of the California Institute of Technology, and the National Science Foundation. Ongoing CARMA development and operations are supported by the National Science Foundation under a cooperative agreement, and by the CARMA partner universities. We have made use of the NASA/IPAC Extragalactic Database (NED), which is operated by the Jet Propulsion Laboratory, California Institute of Technology, under contract with the National Aeronautics and Space Administration. This research has made use of NASA's Astrophysics Data System (ADS).

\end{document}